# Cadmium Zinc Telluride Imager onboard *AstroSat* : a multi-faceted hard X-ray instrument

**A. R. Rao[1], D. Bhattacharya[2], V. B. Bhalerao[2], S.V. Vadawale[3], & S. Sreekumar[4]**

[1]Tata Institute of Fundamental Research, Homi Bhabha Road, Mumbai, India
[2]Inter-University Centre for Astronomy & Astrophysics, Pune, India
[3]Physical Research Laboratory, Ahmedabad, India
[4]Vikram Sarabhai Space Centre, Thiruvananthapuram, India.

**The *AstroSat* satellite is designed to make multi-waveband observations of astronomical sources and the Cadmium Zinc Telluride Imager (CZTI) instrument of *AstroSat* covers the hard X-ray band. CZTI has a large area position sensitive hard X-ray detector equipped with a Coded Aperture Mask, thus enabling simultaneous background measurement. Ability to record simultaneous detection of ionizing interactions in multiple detector elements is a special feature of the instrument and this is exploited to provide polarization information in the 100 – 380 keV region. CZTI provides sensitive spectroscopic measurements in the 20 – 100 keV region, and acts as an all sky hard X-ray monitor and polarimeter above 100 keV. During the first year of operation, CZTI has recorded several gamma-ray bursts, measured the phase resolved hard X-ray polarization of the Crab pulsar, and the hard X-ray spectra of many bright Galactic X-ray binaries. The excellent timing capability of the instrument has been demonstrated with simultaneous observation of the Crab pulsar with radio telescopes like GMRT and Ooty radio telescope.**

**Key Words:** *Gamma-ray bursts, X-ray polarisation, Crab Nebula, X-ray Astronomy*

The prime emphasis of the *AstroSat* satellite is the X-ray timing instrument LAXPC coupled with a sensitive wide field Ultra-violet Imaging Telescope, UVIT[1]. The LAXPC instrument is expected to extend the rich legacy of X-ray timing measurement pioneered by the Rossi X-ray Timing Explorer (RXTE) in an improved manner due to the phenomenally large effective area in the hard X-ray (20 – 80 keV) band. It was, however, realized, from the RXTE experience, that a simultaneous sensitive wide band X-ray spectroscopic measurement would prove extremely crucial to understand the timing behavior of many astrophysical sources. A sensitive soft X-ray telescope (SXT) to cover the lower energy part of the spectrum and CZTI to extend the bandwidth above 80 keV, were included in the *AstroSat* configuration.

CZTI uses the large band gap semiconductor device – Cadmium Zinc Telluride (CZT), which can be operated at near-room temperatures. CZT provides high stopping power, low thermal noise, very good energy and spatial resolutions. CZTI has a geometric area of 976 $cm^2$ and it is achieved by using 64 detector modules divided into four quadrants, each quadrant containing a 4 x 4 matrix of detector modules mounted on a special thermal conductive board (Fig 1, Left). Each module consists of 256 pixelated contacts arranged in a 16x16 array and a digital read out system. The readout includes the energy of incident photon and the address of the pixel where it was measured. CZTI also consists of a veto detector for background rejection and an alpha tagged source for in-orbit calibration. Coarse imaging is achieved using a Coded Aperture Mask which consists of closed and open patterns of squares/rectangles matching the size of the detector pixels. The patterns are based on 255-element pseudo-noise Hadamard Set Uniformly Redundant Arrays. The assembled CZTI in the vibration table is shown in Fig 1, right. The configuration of CZTI is discussed in detail in Bhalerao et al[2]. CZTI has an angular resolution of 17' in the field of view of 4°.6 X 4°.6 (FWHM) and an energy resolution of 6.5 keV (~11% at 60 keV).

The primary scientific objective of CZT Imager is to measure the hard X-ray spectrum of bright X-ray sources in the energy range of 20 – 100 keV. Crab Nebula is used as a spectral calibrator and it is found that a canonical power-law could be fit from 20 keV to 150 keV. The cross-calibration of CZTI with the other AstroSat instruments is going on and in this work we discuss the capability of CZTI as a transient monitor and the timing and polarization features.

## CZTI as a transient monitor

The coded aperture mask and other support structures of CZT Imager have been designed for this energy range and thus become increasingly transparent at energies above 100 keV. However, the 5 mm thick CZT detectors do have significant detection efficiency up to about 400 keV. This gives rise to a unique capability of CZT Imager as a transient monitor at energies above 100 keV. This enables CZT Imager to act as wide field monitor for gamma-ray bursts (GRB). The two layer detector system of CZT Imager consisting of the primary detector plane of CZT detectors along with the veto detector also provides wide band measurements of GRB spectra in the extended energy range of 100 – 800 keV. By extending the technique of the coded mask imaging to the shadows generated by CZTI support structure as well as satellite support structure, CZTI also provides modest localization capability for GRBs.

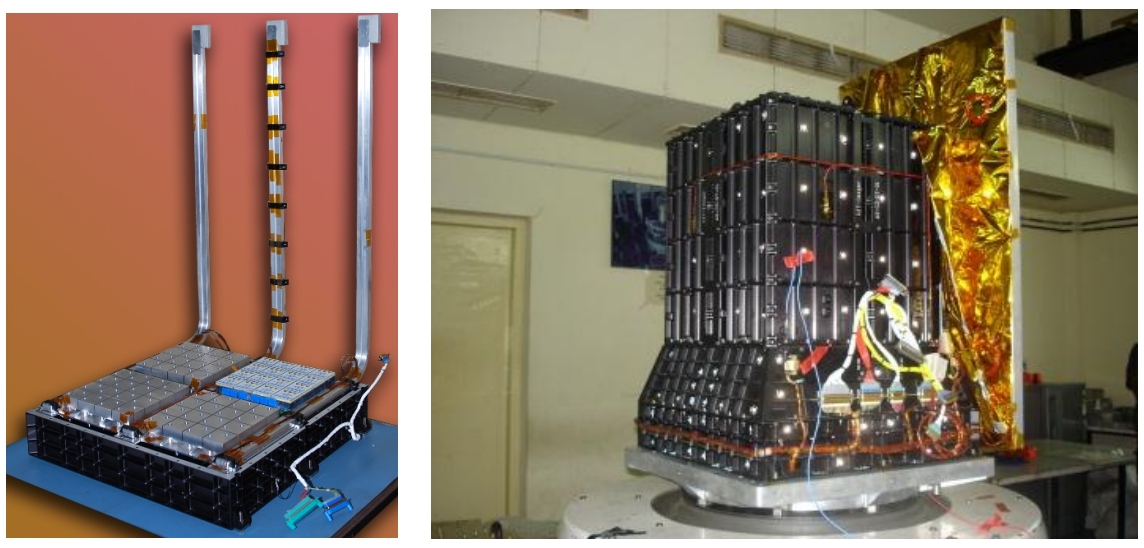

**Figure 1:** ***Left***: The 64 detector modules are assembled in four quadrants, along with the heat pipes for efficient cooling. ***Right***: The assembled CZTI instrument on the vibration table.

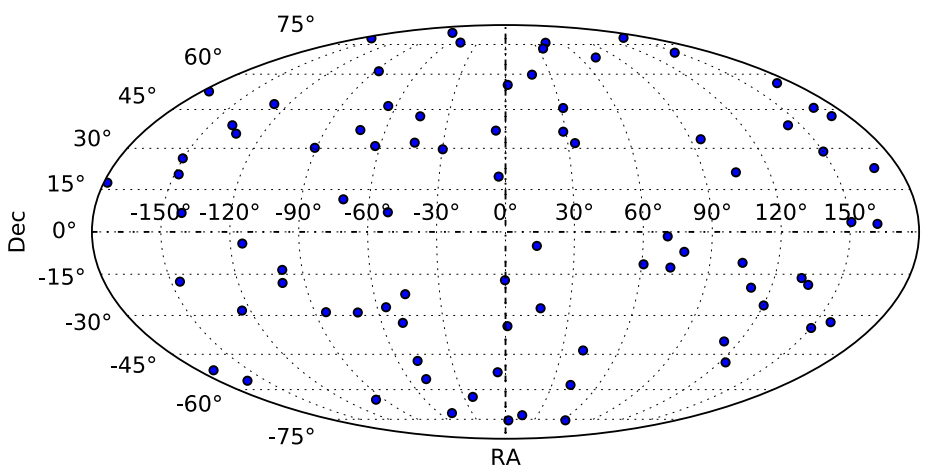

**Figure 2:** GRBs detected by *AstroSat* CZTI from 10 October 2015 to 27 January 2017.

The hard X-ray monitoring capability of CZTI has been well established with in-flight operations. The CZTI was the first scientific instrument to be switched on after the launch of *AstroSat*, and on the first day of science observation CZTI detected GRB 151006A[3,4]. CZTI has now detected over 96 GRBs in sixteen months of operations (Fig 2). Detected GRBs are published as GCN (Gamma Ray Coordinate Network) circulars, and also published online at http://astrosat.iucaa.in/czti/?q=grb.

Another important application of the transient detection capability is the search for electromagnetic counterparts of gravitational wave sources. CZTI has put competitive limits on X-ray emission from GW151226[5], and will continue to be used for X-ray counterpart search, localization and analysis along with current and future gravitational wave observations.

The satellite and CZTI structure are not uniformly transparent to high energy X-rays, and partially obscure sources in different directions. This creates distinct direction-dependent shadows on the detectors. By analyzing these shadow patterns, one can effectively use the entire satellite as a coded aperture for localizing the sources on the sky. This has been demonstrated for two sources so far – by verifying the position of GRB151006A[4], and independently localizing GRB170105A on the sky[6]. For such sources, scattering from different elements of the satellite also becomes an important factor. Hence, detailed simulations of the satellite in GEANT4 have been carried out to calculate expected responses for transients in different directions. Apart from source localization, results from these simulations are also used to calculate spectra of the transient sources.

**Hard X-ray Polarisation**

Another very fascinating capability of CZTI is to measure polarization of incident X-rays in the energy range of 100 – 380 keV. This arises from the pixilated nature of the CZTI detector plane and the significant detection efficiency of the 5 mm CZT thick detectors beyond the primary spectroscopic energy range of CZTI. At energies beyond 100 keV, the incident X-ray photons primarily interact by means of Compton scattering. The Compton scattering is intrinsically sensitive to the polarization of the incident photon in the sense that the incident photon is preferentially scattered in the direction perpendicular to that of the electric field vector of the incident photon. Thus any pixilated detector plane in principle can measure polarization of the incident X-rays. However, in practice it is necessary to have a good combination of various important characteristics such as good efficiency for Compton scattering, appropriate scattering geometry to achieve good modulation factor, capability of processing electronics to preserve the polarimetric information, etc. The design of CZTI achieves an optimal combination of these characteristics to realize scientifically meaningful polarimetric sensitivity for bright X-ray sources.

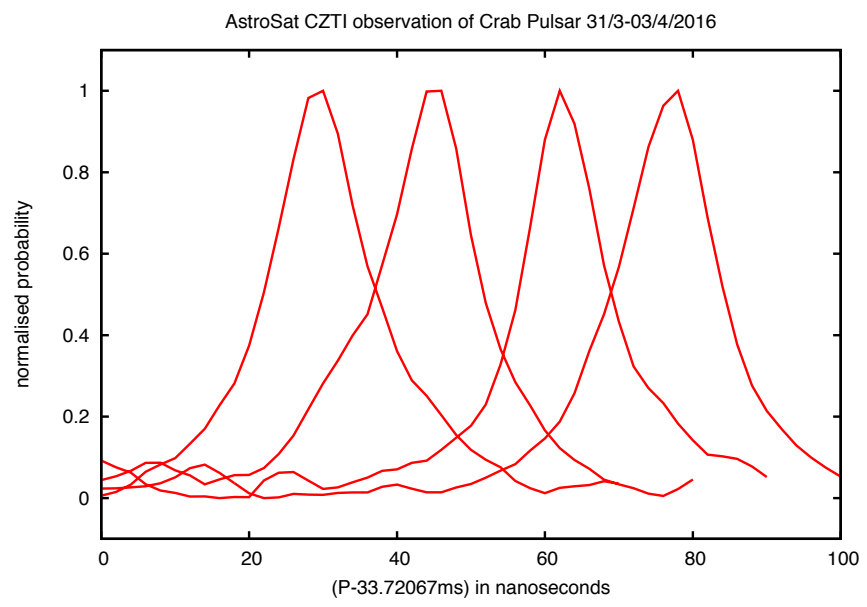

**Figure 3:** Pulse period of the Crab Pulsar estimated from four consecutive sections of a CZTI observation, the length of each section being about half a day. The leftmost curve corresponds to the first section of the data and subsequent sections move progressively to the right. The vertical axis displays normalized fit probability. The best fit period for each section is at the peak of the corresponding curve. The spin-down of the pulsar is clearly demonstrated.

The hard X-ray polarimetric capability of CZTI was extensively investigated by Chattopadhyay et al.[7], using Monte Carlo simulations. It was also experimentally verified before launch using polarized X-rays, and more importantly using un-polarized X-rays[8]. Thus CZTI is the first hard X-ray polarimeter in recent times having ground calibration with un-polarized X-rays. The polarimetric capabilities of CZTI have been well established with observations of the Crab nebula during the first year of *AstroSat* operation. CZTI has provided the first accurate polarization measurement of the Crab nebula in the energy range of 100 – 380 keV with high detection significance. The black hole binary Cygnus X-1 is another bright X-ray source which has shown very interesting indications of polarization signature. However, the most interesting result of the two enhanced capabilities of CZTI i.e. hard X-ray monitoring and hard X-ray polarimetry, is the polarization measurement of GRBs. The number of GRBs detected by CZTI during the initial period are suitable for polarization analysis and a few of them have shown very promising results, including the detection of very high degree of polarization (~90%).

**Timing the X-ray Pulsar in Crab**

The time-tagged event information recorded by the CZTI for every photon detected by it makes it ideal for investigating fast time variability of intensity in hard X-ray sources. One of the best studied sources with fast hard X-ray intensity variation is the Pulsar in the Crab Nebula (see Buhler and Blandford[9] for a review). This is a strongly magnetized neutron star spinning on its axis 30 times a second, so the radiation observed from it has a periodicity of 33 milliseconds. Accurate information about any small change in its spin period is available from continuous monitoring carried out by several radio observatories across the world. As a result, this pulsar has been chosen as the primary timing calibrator of the CZTI instrument and several observations of the Crab have been carried out since the launch of *AstroSat*.

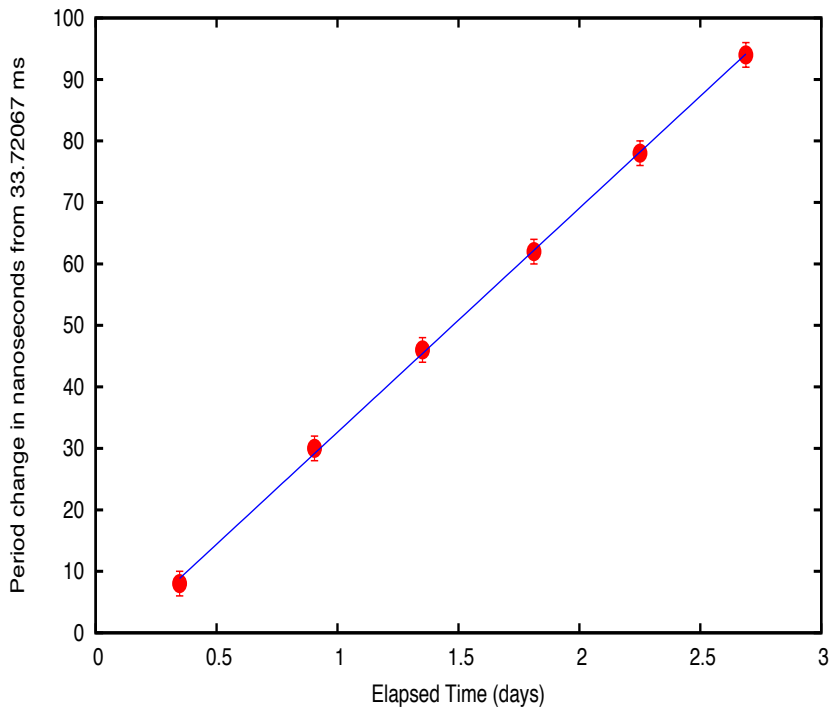

**Figure 4:** The spin-down rate of the Crab Pulsar as measured by *AstroSat* CZTI. The period is determined in six contiguous sections of about half a day each (points). The linear fit plotted as the blue line has a slope of 36.3 nanoseconds per day.

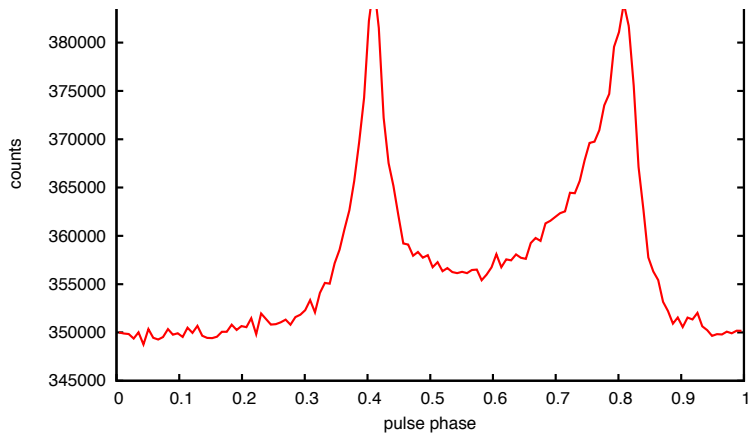

**Figure 5:** Folded pulse profile of the Crab Pulsar in the CZTI energy band of 30 to 250 keV for an observation carried out during March 31-April 3 2016. The time of arrival of the left peak has been used to calibrate the absolute timing of the CZTI instrument.

The event time stamps provided by the local CZTI clock are converted to UTC via correlation with GPS time recorded by a Spacecraft Positioning System on board *AstroSat*. For the timing analysis of the pulsar, the UTC time stamp of each event was used to compute the time of arrival of the corresponding photon at the Solar System Barycentre (SSB), by correcting for the time of flight of photons from the SSB to the moving spacecraft.

An illustration of the timing stability and accuracy of the CZTI is provided by determining the pulse period of the Crab pulsar from different sections of the data, and thereby tracking the evolution of the spin of the pulsar. Figure 3 shows the determination of the best fit pulse period from data sections spaced approximately by half a day. The steady spin-down of the pulsar is immediately apparent. A linear fit to the measured period versus time over a period of about 3 days (Figure 4) yields an estimated spin-down rate of 36.3±0.2 nanoseconds/day, which matches the spin-down rate measured independently by radio observations.

An exercise to calibrate the absolute time reference of the CZTI was then taken up. The data were folded by taking into account the spin period and its derivatives to create a pulse profile from each observation. One such pulse profile is displayed in Figure 5. The absolute time of arrival of the main pulse (first peak) according to the CZTI clock was compared with that of the radio pulse, as estimated by the Ooty Radio Telescope, the Giant Metrewave Radio Telescope and the Jodrell Bank observatory (Avishek Basu et al, in preparation). Ten such determinations using observations spread over the first four months after launch have demonstrated that the CZTI clock has a fixed offset of 2.0±0.5 ms with respect to the Jodrell Bank reference[10] (see also http://www.jb.man.ac.uk/~pulsar/crab.html).

**ACKNOWLEDGMENTS:**

This publication uses the data from the *AstroSat* mission of the Indian Space Research Organisation (ISRO), archived at the Indian Space Science Data Centre (ISSDC). CZT-Imager is built by a consortium of Institutes across India including Tata Institute of Fundamental Research, Mumbai, Vikram Sarabhai Space Centre, Thiruvananthapuram, ISRO Satellite Centre, Bengaluru, Inter University Centre for Astronomy and Astrophysics, Pune, Physical Research Laboratory, Ahmedabad, Space Application Centre, Ahmedabad: contributions from the vast technical team from all these institutes are gratefully acknowledged.